\documentclass{article}

\usepackage{epsfig}
\usepackage{subfigure}
\usepackage{wrapfig}

\usepackage{amsmath}
\usepackage{amssymb}
\usepackage{cite}
\usepackage{array}
\newcommand{\beqs}{\begin{equation*}}
\newcommand{\beq}{\begin{equation}}

\newcommand{\eeqs}{\end{equation*}}
\newcommand{\eeq}{\end{equation}}

\newcommand{\beqas}{\begin{eqnarray*}}
\newcommand{\beqa}{\begin{eqnarray}}

\newcommand{\eeqas}{\end{eqnarray*}}
\newcommand{\eeqa}{\end{eqnarray}}

\newcommand{\eq}[2]{\begin{equation} #1 \label{#2} \end{equation}}

\newcommand{\eqa}[2]{\begin{eqnarray} #1 \label{#2} \end{eqnarray}}

\newcommand{\meq}[2]{\begin{multline} #1 \label{#2} \end{multline}}

\newcommand{\eps}{\varepsilon}
\newcommand{\al}{\alpha}
\newcommand{\be}{\beta}
\newcommand{\ga}{\gamma}
\newcommand{\de}{\delta}
\newcommand{\om}{\omega}
\newcommand{\ka}{\kappa}
\newcommand{\la}{\lambda}
\newcommand{\si}{\sigma}

\newcommand{\Om}{\Omega}

\newcommand{\Si}{\Sigma}

\newcommand{\blist}{\begin{itemize}}

\newcommand{\elist}{\end{itemize}}

\providecommand{\href}[2]{#2}

\begin{document}

\begin{titlepage}

\renewcommand{\thefootnote}{\fnsymbol{footnote}}

\hfill TUW--01--31

\hfill Vers. 1.2 \\

\begin{center}
\vspace{0.5cm}

{\Large\bf Virtual black hole phenomenology from $\boldsymbol{2d}$ dilaton 
theories}
\vspace{1.0cm}
% \vfill

{\bf D.\ Grumiller\footnotemark[1] 
%% and  
%%W. Kummer\footnotemark[2]
}
\vspace{7ex}

{Institut f\"ur
    Theoretische Physik \\ Technische Universit\"at Wien \\ Wiedner
    Hauptstr.  8--10, A-1040 Wien \\ Austria}
%% \vspace{2ex}
\vspace{1.5cm}

\footnotetext[1]{e-mail: {\tt grumil@hep.itp.tuwien.ac.at}}
%%\footnotetext[2]{e-mail: {\tt wkummer@tph.tuwien.ac.at}}

\end{center}

\begin{abstract}

%Equipped with the tools of spherically reduced (dilaton) gravity in first 
%order formulation we discuss the virtual black hole geometry 
%encountered in gravitational scattering of massless $s$-wave scalars. Using 
%the Kerr-Schild decomposition we find a localization of the Ricci scalar with 
%distributional contributions $\de$ and $\de'$. We compare our notion of 
%``virtual black holes'' with Hawking's bubble definition and point out some 
%physical differences.

%Finally we focus on the lowest order $S$-matrix for this model obtained in a 
%previous work (P. Fischer, D. Grumiller, W. Kummer, and D. Vassilevich, 
%{\em Phys. Lett.} {\bf B521} (2001) 357-363). We find indications of (tree 
%level) CPT invariance of our cross-section. As a new feature we observe 
%kinematic pseudo-self-similarity.

Equipped with the tools of (spherically reduced) dilaton gravity in first order
formulation and with the results for the lowest order $S$-matrix for $s$-wave
gravitational scattering (P. Fischer, D. Grumiller, W. Kummer, and D. 
Vassilevich, {\em Phys. Lett.} {\bf B521} (2001) 357-363) new properties of 
the ensuing cross-section are discussed. We find $CPT$ invariance, despite of 
the non-local nature of our effective theory and discover
pseudo-self-similarity in its kinematic sector.

After presenting the Carter-Penrose diagram for the corresponding virtual 
black hole geometry we encounter distributional contributions to its 
Ricci-scalar and a vanishing Einstein-Hilbert action for that configuration. 
Finally, a comparison is done between our (Minkowskian) virtual black hole and 
Hawking's (Euclidean) virtual black hole bubbles.

\end{abstract}

PACS numbers: 04.60.Kz; 04.60.Gw; 11.10.Lm; 11.80.Et

\vfill
\end{titlepage}

\section{Introduction}

Ever since John Wheelers proposal of ``space-time foam'' 
\cite{wheelerrelativity} physicists have toyed with the idea of quantum 
induced topology fluctuations. This has 
culminated not only in the quite successful spin-foam models, which are 
considered as a serious candidate for quantum gravity (cf. e.g. 
\cite{Rovelli:1998yv}), but (among others) also in Hawking's bubble approach 
of virtual black holes (VBH) \cite{Hawking:1996ag}. What are the observational 
consequences of VBHs? 
There have been suggestions that VBHs might lead to loss of quantum coherence 
\cite{Hawking:1997ia}, violation of quantum mechanics \cite{Ellis:1984jz}, 
possibly affecting neutrino-oscillations \cite{Benatti:2001fa}, CPT-violation 
\cite{Huet:1995kr}, non-standard Kaon-dynamics \cite{Benatti:1998vu} etc.

In a recent work \cite{Fischer:2001vz} we drew attention to a purely 
gravitational effect of VBHs, i.e. no other interactions were involved. We 
discussed gravitationally interacting massless scalars for the important 
special case of spherical symmetry in $d=4$ using the tools of first order 
dilaton gravity in $d=2$. After an exact (path) integration of the geometric 
variables (which had been performed elsewhere \cite{Grumiller:2001ea}) we
encountered the VBH (our notion of VBHs differs from Hawking's definition; but 
as we will point out it is justified to call our objects ``virtual black 
holes'', so we stick to this name and apologize for eventual confusion 
caused by this difference). This is remarkable insofar as Hawking's VBHs do 
not seem to be compatible with the special case of spherical 
symmetry because they can only be created in pairs \cite{Hawking:1996ag}. The 
VBH we were describing exists even for that simple scenario. Using this VBH 
we were able to calculate the lowest order $S$-matrix for $s$-wave 
gravitational scattering. As suggested in \cite{'tHooft:1996tq} black holes 
played indeed an important role as intermediate states for the $S$-matrix.

After recalling briefly the main results of \cite{Fischer:2001vz} 
(in section \ref{se:2}) we discuss in the present work new properties of the 
$S$-matrix for $s$-wave gravitational scattering and the corresponding VBH: 
\blist
\item By a simple calculation we observe (tree-level) $CPT$ invariance, which 
is remarkable insofar, as our effective action has been a non-local one and
thus $CPT$-invariance is not granted by the $CPT$ theorem (section 
\ref{se:2.2}).
\item We discuss the kinematic sector of the cross-section for 
(lowest order) gravitational $s$-wave scattering and observe as a new feature
pseudo-self-similarity (section \ref{se:cr}).
\item Although the VBH geometry is a non-local entity we are able to present an
easy interpretable Carter-Penrose diagram resembling a shock-wave geometry with
evaporating shock. The Ricci-scalar of that geometry contains distributional 
contributions $\de$ and $\de'$ and yields a vanishing Einstein-Hilbert action 
(section \ref{se:2.3} and appendix A).
\item Finally, we compare with Hawking's VBHs and point out some parallels -- 
the main difference is rooted in the Minkowskian signature we are using as 
opposed to Hawking's Euclidean approach (section \ref{se:2.4}).
\elist

%This paper is organized as follows:
%
%In section 2 we review briefly the results of \cite{Fischer:2001vz} and 
%elaborate several new conceptual and technical points, e.g. the geometrical 
%interpretation of the virtual black hole, the behavior under time reversal 
%etc. Moreover, we compare our VBH to Hawking's VBH.
%
%In section 3 new properties of the lowest order scattering amplitude and the
%cross-section -- in particular pseudo-self-similarity in the kinematic 
%sector -- are discussed.
%
%Section 4 concludes with an outlook to related open questions.
%
%Appendix A contains the calculation of the Ricci scalar of the VBH geometry
%revealing distributional contributions to it.

\section{From first order gravity to the $\boldsymbol{S}$-matrix}\label{se:2}

We use the same notation as in \cite{Grumiller:2000ah,Grumiller:2001ea,
Fischer:2001vz,Grumiller:2001pt} and restrict our discussion again to 
spherically reduced gravity (SRG), i.e. $4d$ line elements of the form 
($\al,\be=0,1$)
\eq{
ds^2=g_{\al\be}dx^\al dx^\be-X(x^\al)d^2\Om_{S^2},
}{ds4d}
using the signature $(+,-,-,-)$.
The first order action\footnote{For sake of convenience
we repeat all definitions: the action consists of a scalar field 
$\phi$, the dual basis of 1-forms  $e^{\pm}$ (in light-cone gauge), 
the spin connection $\om^a{}_b=:\eps^a{}_b\om$
(which becomes diagonal in light-cone gauge), the dilaton $X$, the Lagrange 
multipliers for torsion $X^{\pm}$ (also in light cone gauge) and the 
``potential'' ${\cal V}(X^+X^-,X)=-1-\frac{X^+X^-}{2X}$ for SRG. The $\ast$ 
denotes the Hodge dual and the integration is performed on a $2d$ manifold 
with Lorentzian signature. We will refer to the 
(Hodge dual of the) last term in (\ref{dil}) as ``matter Lagrangian''. 
Spherical reduction yields the factor $X$ in it as a remnant of the $4d$ 
measure.}
\meq{
L_{\text{FO}} = - \frac{8\pi}{\ka} \int_{M_2} {\big [} X^+ (d - \om) \wedge 
e^- + X^- (d + \om) \wedge e^+ + X d\wedge \omega \\
- e^- \wedge e^+ {\cal V}(X^+X^-,X) + F(X) d\phi \wedge \ast d\phi 
{\big ]}. 
}{dil}
is (classically) equivalent to the spherically symmetric 
Einstein-massless-Klein-Gordon-model, i.e. it reproduces the spherically 
symmetric Einstein equations and the corresponding massless Klein-Gordon 
equation, provided the coupling function is chosen to be $F(X)=-\ka X/2$ with 
$\ka=8\pi G_N:=1$. Roughly
speaking, it combines the advantages of a Hamiltonian formulation 
(``symplectic'' structure\footnote{Without matter (\ref{dil}) can be shown to 
be a special case of a Poisson-$\si$ model \cite{Schaller:1994es}, i.e. 
strictly speaking there is no symplectic structure but only a Poisson 
structure. Even the matter part can be brought into first order form, but since
the selfdual and anti-selfdual components of the scalar field will mix in 
general this is of minor interest -- cf. e.g. \cite{Strobl:1999wv}.}, first 
order action) with the advantages of a Lagrangian formulation (manifest 
diffeomorphism invariance) \cite{Klosch:1996fi}.

These advantages, combined with a convenient choice of gauge\footnote{For 
historical reasons it has been called ``temporal gauge''. This choice 
fixes $\om_0=0$, $e^-_0=1$ and $e^+_0=0$. It yields a metric in outgoing 
Sachs-Bondi form.}, allow for an exact path integral quantization of the 
geometric part of (\ref{dil}) \cite{Kummer:1997hy,Kummer:1998zs}. However, the 
final matter-integration could only be performed perturbatively (supposing the 
scalar particles have a typical energy which is small compared to the Planck 
energy).

\subsection{Minimal vs. nonminimal coupling}\label{se:2.1}

A comparison between the results for (in $d=2$) minimally coupled scalars 
(i.e. with $F(X) = $ const.) \cite{Grumiller:2000ah} and spherically reduced 
scalars shows not only quantitative but also essential qualitative changes 
\cite{Fischer:2001vz}. Remarkably, these changes can be seen already at a 
rather fundamental level: the constraint algebra of the (first class) 
secondary constraints, $\{G_i(x), G_j(x')\}=\de(x-x')C_{ij}{}^k(x) G_k(x)$, 
with $i=1,2,3$ and $\{\bullet,\bullet\}$ being the Poisson
bracket. The nonvanishing structure functions are given by 
\cite{Grumiller:2001ea}
\newline \parbox{3cm}{\begin{eqnarray*}
&& C_{12}{}^2 = -1, \\
&& C_{13}{}^3 = 1,
\end{eqnarray*}} \hfill
\parbox{7cm}{\begin{eqnarray*}
&& C_{23}{}^1 = -\frac{\partial{\cal{V}}}{\partial X}+\frac{F'(X)}
{(e)F(X)}{\cal{L}}^{(m)}, \\
&& C_{23}{}^2 = -\frac{\partial {\cal{V}}}{\partial X^+}, \\
&& C_{23}{}^3 = -\frac{\partial {\cal{V}}}{\partial X^-}.
\end{eqnarray*}} \hfill
\parbox{1cm}{\begin{equation} \label{structure} \end{equation}} \hfill
\newline
Without matter or in the simple case of $F(X) = $const. the exceptional term 
in (\ref{structure}) proportional to $dF/dX$ vanishes. 
We call this term exceptional because as opposed to all other terms in
(\ref{structure}) it is not only a function of the dilaton $X$ and the 
auxiliary fields $X^\pm$, but it also depends on the scalar field, its 
momentum and the vielbein.
%We call this term exceptional because it destroys the simple finite W-algebra structure that we would have otherwise (for a definition and the relevance of finite W-algebras cf. e.g. \cite{deBoer:1996nu}). 
This qualitative change of the 
constraint algebra has far-reaching technical consequences: the system of 
partial differential equations encountered in the quantization procedure
was uncoupled in the minimally coupled case \cite{Grumiller:2000ah}, but 
becomes coupled for SRG \cite{Grumiller:2001ea}. This coupling turns out to be 
the reason why for the latter case we obtain two vertices to lowest order 
while the former produced only one. Since all of them yield divergent 
contributions to the $S$-matrix the only hope for a non-trivial finite result
would be a complete cancellation of the two divergent contributions in the 
non-minimally coupled case. It has been shown that, indeed, the ``miracle'' 
happens: all divergent contributions cancel, but fortunately some non-trivial 
terms remain \cite{Grumiller:2001ea,fischervertices,Fischer:2001vz}. We 
conjectured this apparent coincidence being a result of gauge independence of 
the $S$-matrix and the occurrence of intermediate divergencies as an artifact 
of our particular gauge. 

\subsection{The scattering amplitude}\label{se:sc}

The result for the four particle scattering amplitude with ingoing scalar 
$s$-wave modes $q,q'$ and outgoing ones $k,k'$ reads \cite{Fischer:2001vz,
Grumiller:2001ea}:
\eq{
T(q, q'; k, k') = -\frac{i\ka\de\left(k+k'-q-q'\right)}{2(4\pi)^4
|kk'qq'|^{3/2}} E^3 \tilde{T}
}{RESULT}
with the total energy $E=q+q'$,
\eqa{
&& \tilde{T} (q, q'; k, k') := \frac{1}{E^3}{\Bigg [}\Pi \ln{\frac{\Pi^2}{E^6}}
+ \frac{1} {\Pi} \sum_{p \in \left\{k,k',q,q'\right\}}p^2 \ln{\frac{p^2}{E^2}} 
\nonumber \\
&& \quad\quad\quad\quad\quad\quad \cdot {\Bigg (}3 kk'qq'-\frac{1}{2}
\sum_{r\neq p} \sum_{s \neq r,p}\left(r^2s^2\right){\Bigg )} {\Bigg ]},
}{feynman}
and the momentum transfer function\footnote{The square of the momentum 
transfer function is similar to the product of the 3 Mandelstam variables 
$stu$ -- thus we would have non-polynomial terms like $\ln{(stu)^{stu}}$ in 
the amplitude, which is an interesting feature. However, the usual Mandelstam 
variables are not available here, since we do {\em not} have momentum 
conservation in our effective theory (there is just one $\de$-function of
energy conservation).} $\Pi = (k+k')(k-q)(k'-q)$. The 
interesting part of the scattering amplitude is encoded in the scale 
independent factor $\tilde{T}$. 

\section{CPT invariance?}\label{se:2.2}

Since we were only able to perform the path integral quantization in temporal 
gauge, the conjecture about gauge independence is hard to prove
explicitly. However, we can at least look at the ingoing Sachs-Bondi 
case ($\om=0$,$e^-_1=-1$,$e_0^+=0$), which induces only minor changes
in the formulae of \cite{Fischer:2001vz}. Tracing through the whole 
calculation yields the same vertices, but with negative overall 
sign\footnote{The fastest way to check this is as follows: $e_0^-$ changes 
sign by assumption, implying an overall sign change of the gauge fixed 
Hamiltonian after integrating out all the ghost terms. One can accommodate these
changes most easily in eq. (6) of \cite{Fischer:2001vz} by changing the sign of
the $p_i\dot{q}_i$ term. Since we have fixed $p_1$ by boundary conditions it
will remain the same, but $p_2$ and $p_3$ will flip their sign. Eq. (10) of 
that work implies no flip for $q_3$, but $q_2$ will change (as expected we 
obtain the line element (\ref{ds}) of the present work but with $K \to -K$ and 
$u\to v$). With the redefinitions $c_1\to-c_1$ and $S_1\to-S_1$ one obtains 
the vertices (19) and (20) of that work but with an overall sign.}. This 
changes also the sign in the final formula for the amplitude. 

Because the latter is purely imaginary we can provide the following 
interpretation: outgoing and ingoing Sachs-Bondi gauges are connected (or 
rather: disconnected) by a time 
reversal transformation $T$. As expected, time reversal acts like complex 
conjugation on the scattering amplitude. The cross section for $s$-waves is 
not affected by this transformation and hence $T$-invariant. Charge conjugation
$C$ acts trivially on our model because we have considered only uncharged
particles. Also the parity transformation $P$ does not induce any changes 
because our amplitude respects spherical symmetry {\em per constructionem}.
Thus, we have $CPT$-invariance. We interpret this as an indication of our 
conjecture's validity. It is an interesting result by itself, since the 
standard proof of $CPT$-invariance needs locality as a requirement
\cite{Streater:1989vi} and our effective action is non-local 
\cite{Fischer:2001vz,Grumiller:2001ea}. It would be an interesting task to 
check whether this $CPT$ invariance is violated by 1-loop effects or not. The 
results obtained in \cite{Huet:1995kr} seem to suggest that such a violation 
will occur.

\section{Kinematic discussion of the cross-section}\label{se:cr}

With the definitions $k=E\al$, $k'=E(1-\al)$, $q=E\be$, and $q'=E(1-\be)$ 
($\al,\be\in[0,1]$, $E\in\mathbb{R}^+$) a cross-section like quantity for 
spherical waves can be defined \cite{fischervertices,Fischer:2001vz}:
\begin{equation}
 \frac{d\sigma}{d\alpha}=\frac{1}{4(4\pi)^3}\frac{\kappa^2 E^2 |\tilde{T}
(\alpha, \beta)|^2}{(1-|2\beta-1|)(1-\alpha)(1-\beta)\alpha\beta}.
\label{crosssection}
\end{equation}
Because this is (at least in principle) a measurable quantity it is of some 
interest to discuss it further. Since a word says less than $10^{-3}$ pictures
we prepared some kinematic plots (the dependence of the cross-section
on the total incoming energy is trivially given by the monomial pre-factor 
$E^2$ -- the cross-section like quantity (\ref{crosssection}) vanishes in the 
IR limit and diverges quadratically in the UV limit; at least the last fact 
is not very surprising: considering our assumption of energies being small as 
compared to Planck energy it simply signals the breakdown of our perturbation
theory).

\begin{figure}
\noindent
\begin{minipage}[b]{\linewidth}
  \centering
  \subfigure[$\al \in \left(0,1\right)$, $\be \in \left(0,1\right)$]
            {\epsfig{file=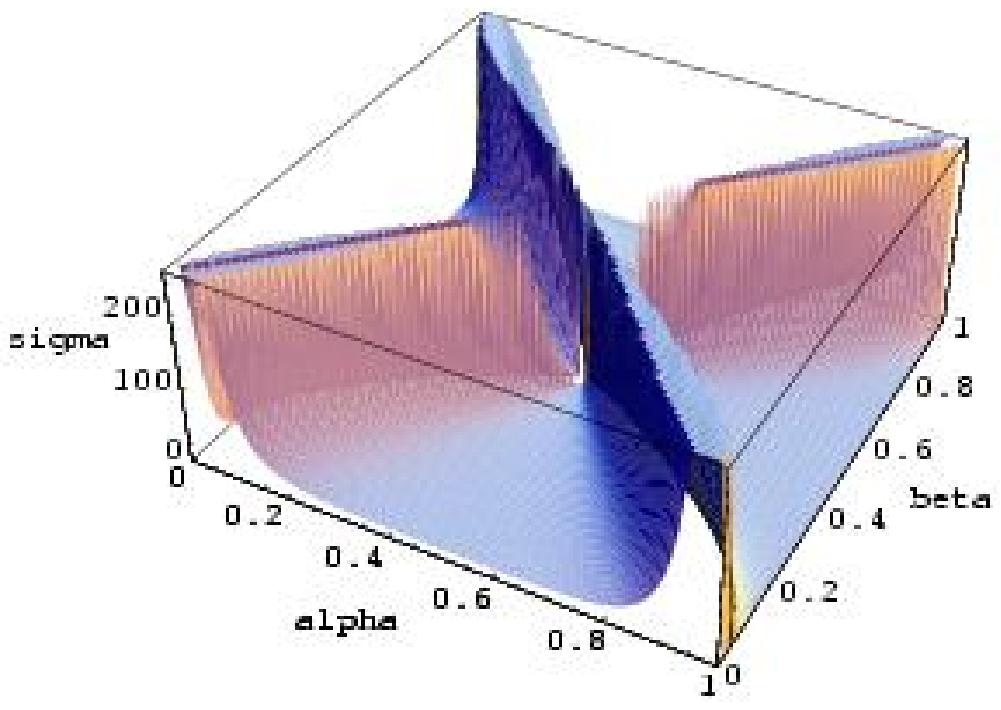,width=.47\linewidth}} \hfill
  \subfigure[$\al \in \left(0,1\right)$, $\be \in \left(0,0.5\right)$] 
            {\epsfig{file=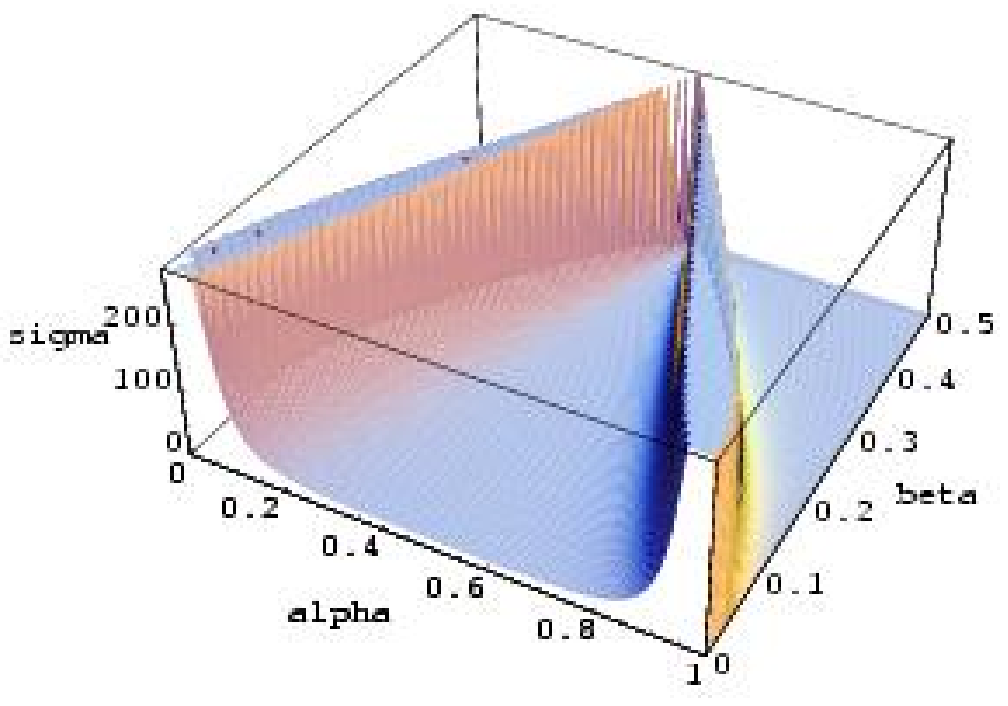,width=.47\linewidth}}  
  \subfigure[$\al \in \left(0,1\right)$, $\be=\al+\eps$ with $\eps \in \left(-0.1,0.1\right)$]
            {\epsfig{file=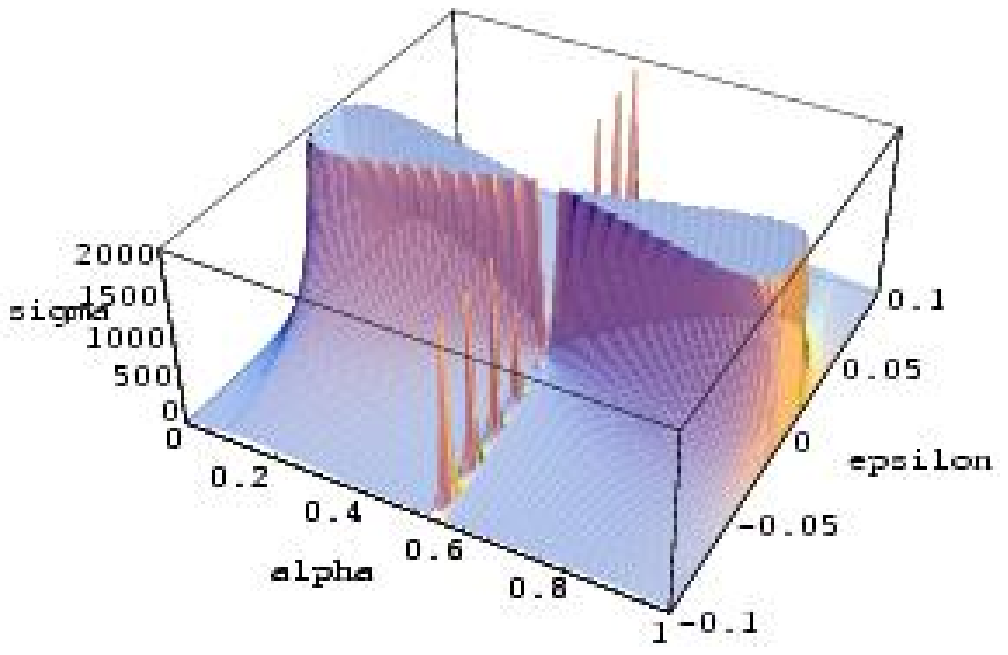,width=.47\linewidth}} \hfill
  \subfigure[$\al \in \left(0.4,0.6\right)$, $\be=\al+\eps$ with $\eps \in \left(-0.1,0.1\right)$]
            {\epsfig{file=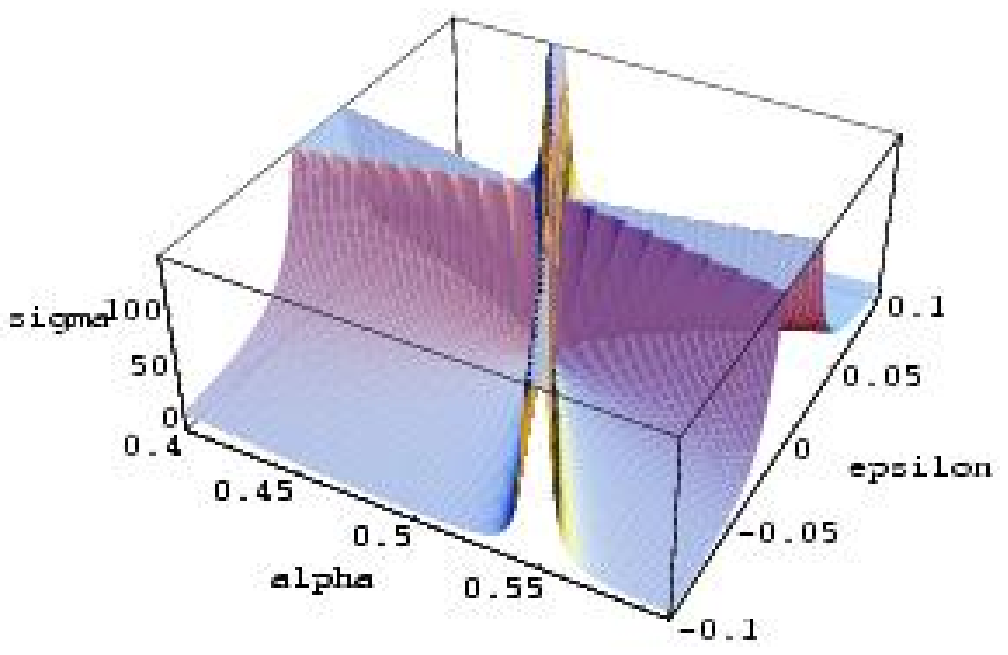,width=.47\linewidth}}
  \subfigure[$\al \in \left(0.4999,0.5001\right)$, $\be \in \left(0.4999,0.5001\right)$]
            {\epsfig{file=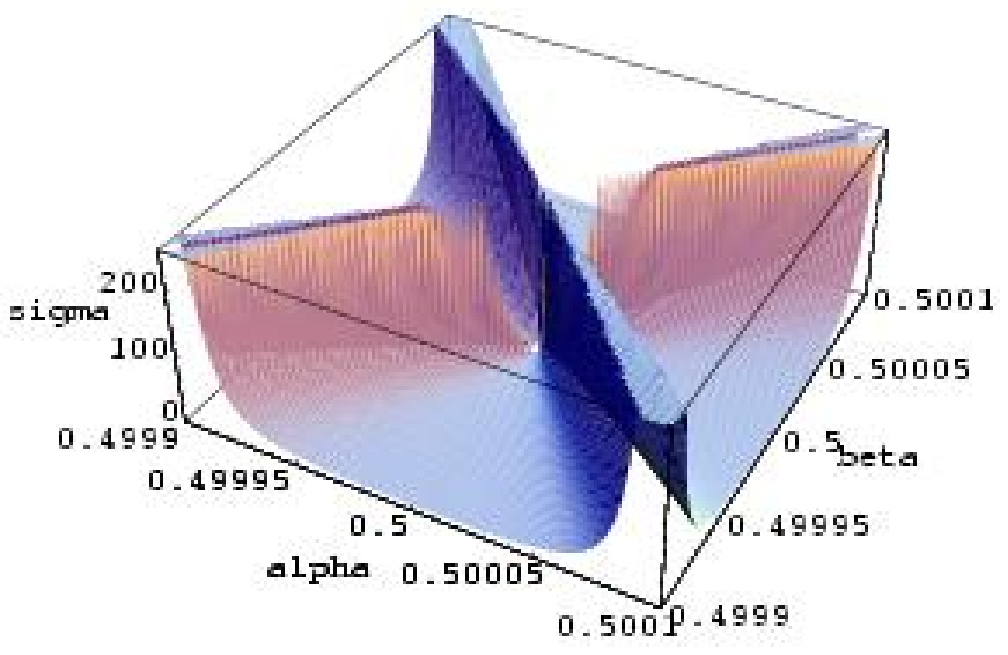,width=.47\linewidth}} \hfill
  \subfigure[$\ga \in \left(0,0.5\right)$, $\de \in \left(0.95,0.99\right)$]
            {\epsfig{file=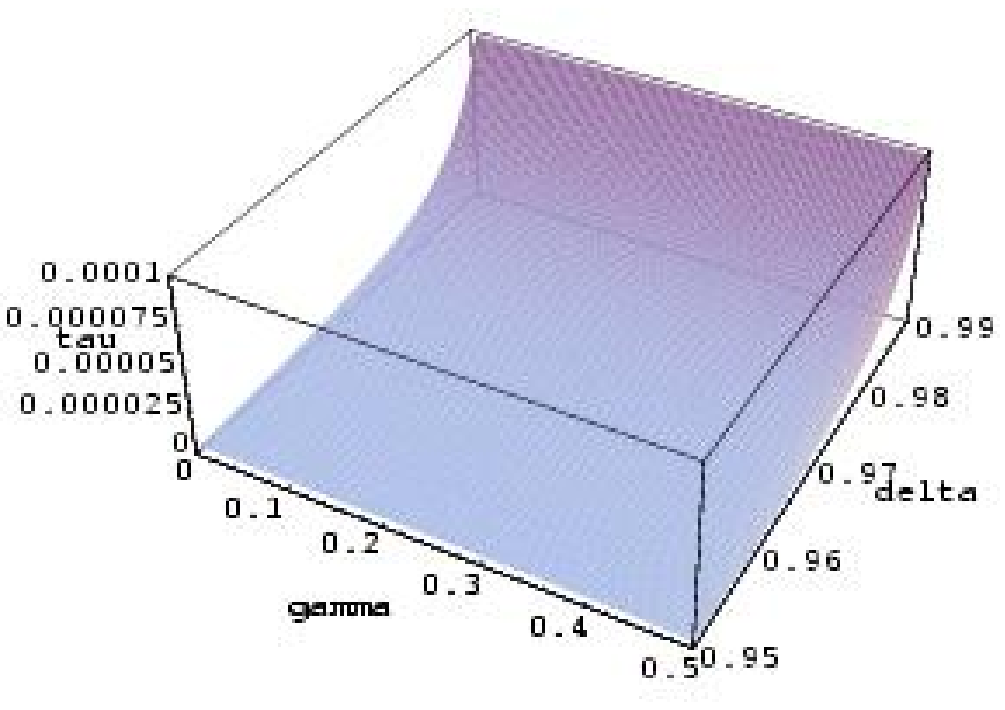,width=.47\linewidth}}
\end{minipage}
\caption{Kinematic plots of the $s$-wave cross-section $d\si/d\al$}
\label{fig:kin}
\end{figure}

The complete kinematic region is plotted in figure \ref{fig:kin}(a).
Figure \ref{fig:kin}(b) shows a kinematic plot for $\al \in [0,1]$ and $\be 
\in [0,1/2]$ (the units of $d\si/d\al$ are irrelevant for our kinematical
discussion). One can easily see the symmetry $\al \leftrightarrow (1-\al)$.
Further, the forward scattering poles $\al=\be$ or $1-\al=\be$ are clearly
visible. Apart from these poles apparently no interesting structure is present
at first glance. Numerical instability of the used {\em Mathematica} 
\cite{wolframmathematica} routines 
leads to ``holes'' in that plot close to the forward scattering 
poles. Figure \ref{fig:kin}(c) zooms into that region with higher resolution.
The interesting looking spoke-like structure corresponds again to forward 
scattering. However, figure \ref{fig:kin}(d) reveals the spokes as a 
numerical artifact (there should be a plateau instead). Plot \ref{fig:kin}(e) 
shows apparently self-similarity of the scattering amplitude: it looks 
identical to \ref{fig:kin} (a), the plot showing the whole kinematic region, 
although the second one is 
zoomed in by a factor of 10000. This kind of self-similarity in the kinematic 
sector is a property of (\ref{crosssection}) which has not been discussed 
previously.

\begin{figure}
\noindent
\begin{minipage}[b]{\linewidth}
  \centering
  \subfigure[$\al \in \left(0,0.9\right), \be=\al+0.1$] 
            {\epsfig{file=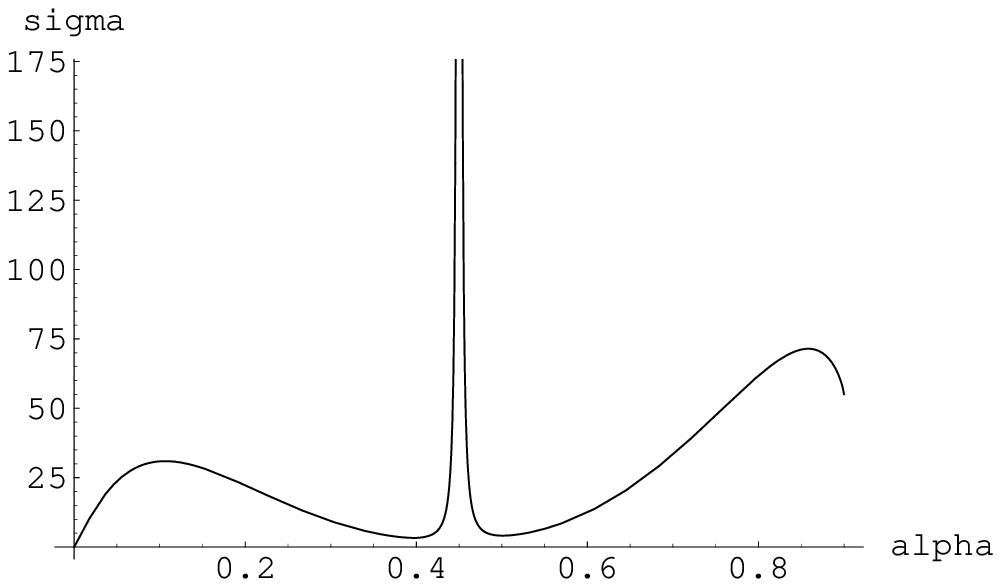,width=.47\linewidth}} \hfill
  \subfigure[$\al \in \left(0.1,1\right), \be=\al-0.1$]
            {\epsfig{file=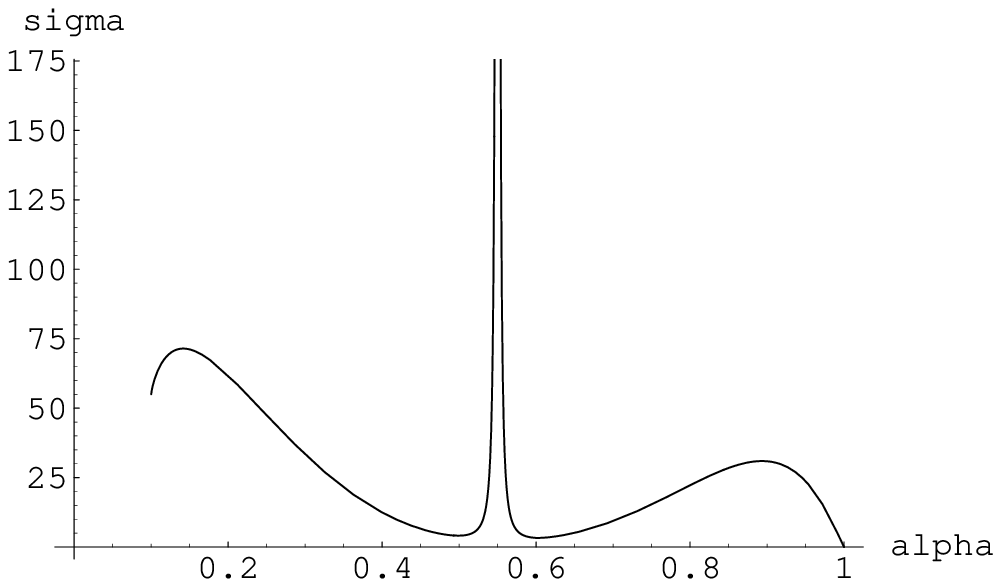,width=.47\linewidth}}
  \subfigure[$\al=0.5, \be \in \left(0,1\right)$]
            {\epsfig{file=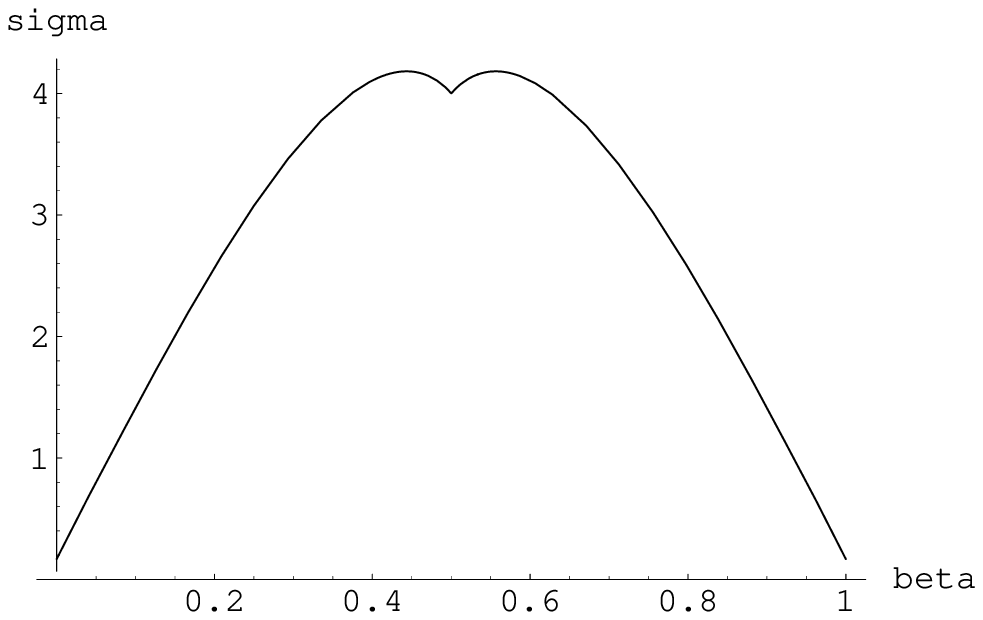,width=.47\linewidth}} \hfill
  \subfigure[$\al \in \left(0,1\right), \be=0.5$]
            {\epsfig{file=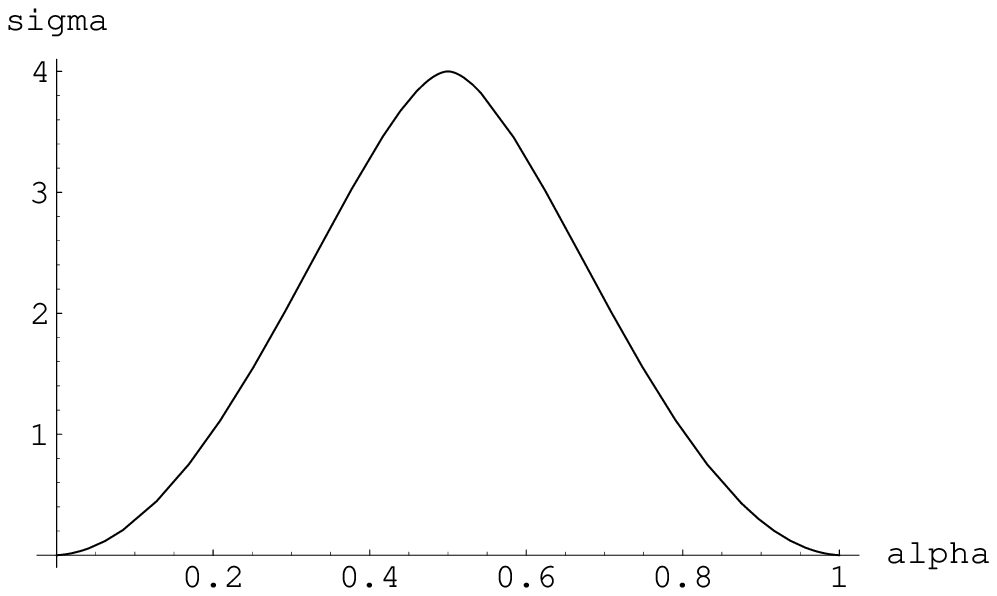,width=.47\linewidth}}
  \subfigure[$\al=0.51, \be \in \left(0,1\right)$]
            {\epsfig{file=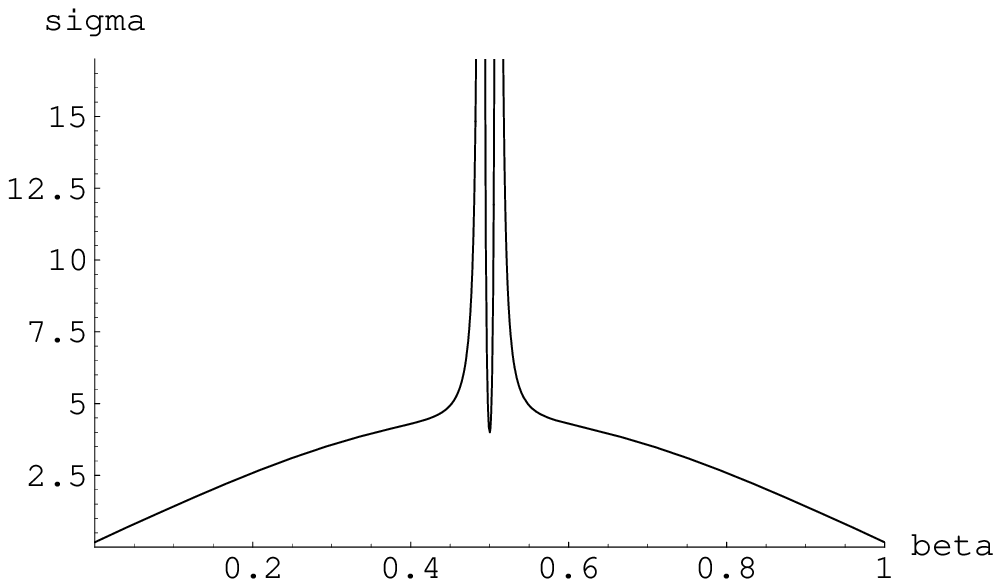,width=.47\linewidth}} \hfill
  \subfigure[$\al=0.5, \be \in \left(0.49996,0.50004\right)$]
            {\epsfig{file=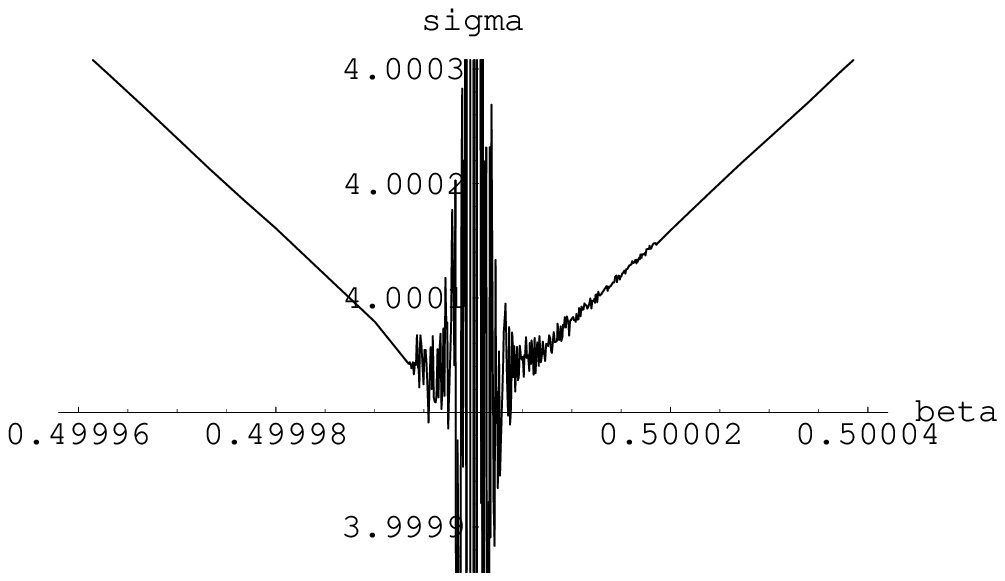,width=.47\linewidth}}
\end{minipage}
\caption{Cuts through the kinematic plots}
\label{fig:cut}
\end{figure}

Before starting such a discussion we consider cuts through the kinematic 
plots, because they reveal a hidden substructure: figure \ref{fig:cut}(a) 
shows apart from the forward scattering pole two local maxima in the 
cross-section. The mirror counterpart is depicted in \ref{fig:cut}(b) -- 
one sees clearly the symmetry between the graphs, a relic of the 
$\al\leftrightarrow (1-\al)$ symmetry. Since the center of the kinematic 
region ($\al\approx 0.5\approx\be$) is not distinguishable very well in the 
previous plots some cuts through that region are helpful. The graphs 
\ref{fig:cut}(c) and \ref{fig:cut}(d) show, that the limiting value 
$\al=\be=0.5$ seems to be finite. However, close to that value the forward 
scattering peaks still exist, as shown in \ref{fig:cut}(e). The final diagram 
\ref{fig:cut}(f) demonstrates the numerical instability close to the center. 
Of course, the spikes are artificial. Plugging the central values 
$\al=\be=0.5$ into (\ref{crosssection}) yields a (pole-like) divergence which 
is not seen in \ref{fig:cut}(c) and \ref{fig:cut}(d), but this is due to the 
numerical instability shown in \ref{fig:cut}(f). So the forward scattering 
pole occurs also in the center of the kinematic region. 

Surprisingly, the feature of self-similarity is absent in the cut graphs of
figure \ref{fig:cut}. To 
resolve this puzzle it is helpful to reparameterize the scattering amplitude 
as a function of $\ga=\al-1/2$ and $\de=(\be-1/2)/(\al-1/2)$ with the 
ranges $\ga\in[-1/2,1/2]$ and $\de\in[-\infty,\infty]$ (due to the symmetries
in (\ref{crosssection}) it suffices to restrict oneself to the ranges 
$\ga\in[0,1/2]$ and $\de\in[0,\infty]$). Self-similarity of the 
cross-section (which now depends on $\ga$ and $\de$) would mean its 
independence of $\ga$ (up to a scale factor). 
However, it is straightforward to check that (\ref{crosssection}) is not 
$\ga$-independent. So why do we seem to observe self-similarity? The answer 
is contained in figure \ref{fig:kin} (f) which plots the function 
$\tau(\ga,\de)$, defined by
\eq{
\tau:=\frac{d\si}{d\al}(\ga,\de)/\frac{d\si}{d\al}(\ga,\de_0).
}{tau}
It is proportional to the cross-section (\ref{crosssection}) but rescaled such 
that at $\de=\de_0:=0.99$ (i.e. very close to the forward scattering peak) its 
value is fixed to 1 for all $\ga$. Apparently, there is no $\ga$ dependence 
visible in that plot, although a (hidden) substructure exists. 

Analytically this can be seen most easily by 
expanding (\ref{crosssection}) close to $\de=1$:
\eq{
\frac{d\si}{d\al}=\frac{1}{\de-1}A(\ga)+B(\ga) + {\cal O}(\de-1),
}{deseries}
with some simple but irrelevant functions $A(\ga)$ and $B(\ga)$. 
Plugging this result into (\ref{tau}) yields
\meq{
\tau\approx\frac{(\de_0-1)}{(\de-1)}\left(1-(\de_0-1)\frac{B}{A}\right)+
(\de_0-1)\frac{B}{A}+{\cal O}((\de_0-1)^2) \\
+{\cal O}((\de-1)(\de_0-1))+{\cal O}((\de_0-1)^3/(\de-1)).
}{detau}
For $\de\approx\de_0$ (this corresponds to the visually essential parts of the 
plots \ref{fig:kin} (a) and (e)) the leading order (LO) is approximately 1 
(and $\ga$ independent) and the NLO vanishes\footnote{The NLO is given by
$\frac{B}{A}(\de_0-1)(1-\frac{\de_0-1}{\de-1})$, but for $\de \approx \de_0$
this is one order smaller and hence part of NNLO.}. The self-similarity is 
broken only on the NNLO level in this kinematic region. Far away from the 
forward scattering peak the quantity $\tau$ is suppressed and deviations from
self-similarity are present but not visible in the plots since they are scaled
with $(\de_0-1)$. Thus, we conclude that no (kinematic) self-similarity exists
in our cross-section (\ref{crosssection}), 
but only a pseudo-self-similarity close to the forward scattering peaks.

As a final {\em caveat} we should emphasize that the full $4d$ theory -- even
in the special case of spherical symmetry -- is more complex than the
spherically reduced model we have treated. This is of particular importance at
the quantum level \cite{Frolov:1999an,Sutton:2000gm}. 

\section{VBH geometry}\label{se:2.3}

\begin{wrapfigure}{r}{3cm}
\epsfig{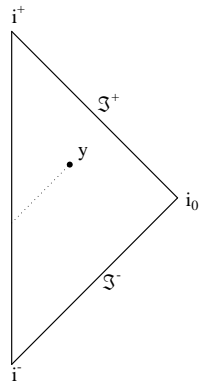}
\caption{CP diagram of the VBH}
\label{fig:cp}
\end{wrapfigure}

Before we compare our VBH with Hawking's VBH we would like to shed some light 
on the geometric role of its effective line element \cite{Fischer:2001vz}: 
\begin{equation}
(ds)^2=2drdu+\left(1-\frac{2m(r,u)}{r}-a(r,u)r\right)(du)^2,
\label{ds}
\end{equation}
with $m(r,u):=\delta(u-u_0)\theta(r_0-r)m_0(r_0)$ and 
$a(r,u):=\delta(u-u_0)\theta(r_0-r)a_0(r_0)$. The $\theta$-function is 
responsible for a discontinuity: for $r\to\infty$ neither the Schwarz\-schild 
term proportional to $m$ nor the Rindler term proportional to $a$ is present.
The $\de$-function restricts the non-vacuum part to the outgoing null line
depicted in figure \ref{fig:cp} (the time reversed VBH is localized on an 
ingoing null line). The functions $m_0$ and $a_0$ are given 
by\footnote{The constants $d_0,d_1$ are given by $d_0:=-c_0/2^{5/2}$, 
$d_1:=-c_1/2^{7/2}$. The quantities $c_0,c_1$ have been defined in 
\cite{Fischer:2001vz}.}
\eq{
m_0:=d_1 r_0^2 + d_0 r_0^3, \hspace{0.5cm}a_0:=-2d_1 + 3d_0 r_0.
}{m0a0}
It is amusing, that the ``mass'' $m_0$ does not only contain a
``volume'' term ($d_0r_0^3$) but also a ``surface'' term ($d_1r_0^2$).

This line element was obtained in part by fixing the boundary values of
the geometric quantities (thus our effective geometry is asymptotically flat 
{\em per constructionem}) and in part by reconstructing geometry out of the
matter field (this was possible because we have treated the geometric part 
exactly; for details we refer to \cite{Fischer:2001vz,Grumiller:2001ea}). The 
corresponding Carter-Penrose (CP) diagram is presented in figure \ref{fig:cp}.

It needs some explanations: first of all, the effective line element is 
non-local in the sense that it depends not only on one set of coordinates (e.g.
${u, r}$) but on two ($x=(u,r), y=(u_0,r_0)$). This non-locality was a
consequence of integrating out geometry non-perturbatively 
\cite{Grumiller:2001ea}. For each selection of $y$ 
it is possible to draw an ordinary CP-diagram treating $u_0, r_0$ as external 
parameters. The light-like ``cut'' in figure \ref{fig:cp} corresponds to 
$u=u_0$ and the endpoint labeled by $y$ to the point $x=y$. The 
non-trivial part of our effective geometry is concentrated on the cut.

Of course, one
should not look at the line elements in given coordinates, but at the
scalar invariants of the Riemann tensor in order to obtain physical insight.
For SRG it suffices to consider the Ricci-scalar, since there is only one 
independent physical quantity in the Riemann tensor. 
As inspection of the equations of 
motion for the spin connection immediately reveals (\ref{eom3}), it has a 
distributional concentration at the endpoints of the cut and ``propagates'' 
along the light-like cut. This shows that the appearance of such a cut 
is not an artifact of the chosen coordinates, but has a physical counterpart 
in the curvature scalar (for details of that calculation we refer to 
appendix A): 
\meq{
R^{VBH} (u,r;u_0,r_0) = \de(u-u_0){\Bigg [}-\de'(r-r_0)\left(\frac{2m_0}{r}
+a_0r\right) \\
-\de(r-r_0)\left(\frac{4m_0}{r^2}+6a_0\right) +\Theta(r_0-r)\frac{6a_0}{r} + 
2\frac{m_0}{r^2}\de(r){\Bigg ]}.
}{ricciscalar}
The last contribution corresponds to the Schwarzschild singularity and has 
been advocated in \cite{Balasin:1994kf}. It is quite interesting to observe
what happens when plugging (\ref{ricciscalar}) into the Einstein-Hilbert
action ($\int_0^\infty dr\de(r):=1$):
\meq{
L^{VBH}_{EH}=\left.\int_{M_4}d^4x\sqrt{-g^{(4)}}R\right|_{VBH}=4\pi
\int\limits_{-\infty}^{\infty} du \int\limits_0^{\infty} r^2 dr 
R^{VBH}(u,r;u_0,r_0) \\ 
= 4\pi(2m_0+3a_0r_0^2-4m_0-6a_0r_0^2+3a_0r_0^2+2m_0) = 0.
}{EH}
It vanishes.

Apart from the cut, the spacetime coincides with (Minkowski) vacuum. It is fair
to ask for precedents of such a geometry: on the one hand, our VBH seems 
similar to shock wave geometries (cf. e.g. \cite{Aichelburg:1971dh}), but in
our case the ``shock'' suddenly ``evaporates'' at $x=y$. In the matter sector,
the VBH naturally shows resemblance to the quantum black hole obtained in the 
thin shell approach (cf. e.g. \cite{Berezin:1998fn}) since in both cases the
(spherically symmetric) shells are localized in space, but in our case the
(off-shell) scalar field is additionally localized in time (cf. eq. (8) and 
(9) in \cite{Fischer:2001vz}). On the other hand, -- on a superficial level -- 
our effective geometry could be interpreted as a soliton or an instanton 
solution since the non-vacuum part is ``localized'', but as opposed to a 
soliton (localized on (a compact family of) time-like trajectories) or an 
instanton (localized in (Euclidean) space-time) we have a light-like 
localization, a ``nulliton'', existing only for a finite affine parameter 
(which is roughly speaking the light-like ``equivalent'' of proper time) 
until it ``evaporates'' (thus, it is not stable as opposed to solitons). 
Moreover, our geometry is singular at $x=y$ and soliton solutions usually are 
required to be regular. The singularity appears to be an inevitable 
consequence of the light-like structure\footnote{To elucidate this point, let 
us discuss the Klein-Gordon equation with arbitrary potential $f(\phi)$ on a 
flat $2d$ Minkowski background in light-cone gauge: 
$\partial_+\partial_-\phi+f'(\phi)=0$. We assume the solution to be outgoing 
(or, in the complexified version 
$\partial_{\bar{z}}\partial_z \phi(z,\bar{z})+f'(\phi)=0$ the field must be 
(anti)holomorphic). A sufficient condition for this property is 
$\partial_-\phi=0$. Then the second order partial differential equation 
reduces to an ordinary equation $f'(\phi)=0$. The set of solutions to 
this equation is given by the local extrema of $f(\phi)$, i.e. each solution 
yields a constant function (e.g. $\phi=\{0,\pm m/\sqrt{\la}\}$ for the Higgs 
potential $f(\phi)=-m^2\phi^2/2+\la \phi^4/4$). These are the trivial (vacuum) 
solutions. However, if we allow for distributional source terms in the 
potential non-trivial holomorphic solutions can exist. The simplest example is 
$\phi=\ln{z}$. Since $\partial_{\bar{z}}1/z=2\pi\de^{(2)}(z,\bar{z})$ (cf. 
e.g. \cite{Polchinski:1998rq}) the solution can be non-constant only if the 
potential contains a distribution (``source term''). Remarkably, our VBH has 
similar properties: it is null, outgoing, localized in the outgoing null 
direction and has a $\de$-contribution.}. So it seems this geometry is without 
precedent.

We do not want to suggest to take the effective geometry at face value -- this 
would be like over-interpreting the role of virtual particles in a loop 
diagram. It is a nonlocal entity and we still have to ``sum'' (read: 
integrate) over all possible geometries of this type in order to obtain the 
nonlocal vertices and the scattering amplitude. But nonetheless, the 
simplicity of the effective geometry, the fact that one has to sum 
over all possible configurations and the vanishing of the Einstein-Hilbert 
action are nice qualitative features of this picture. 

\section{Comparison with Hawking's VBH}\label{se:2.4}

We should clarify the difference between ``our'' VBH and ``Hawking's''
VBH: while the latter is defined by its topological structure ($S^2\times 
S^2$) in Euclidean space \cite{Hawking:1996ag} our definition relies on a 
discontinuity in a ``phenomenological'' quantity, the geometric part of the 
conserved quantity existing in all $2d$ dilaton theories \cite{Kummer:1992rt,
Grosse:1992vc,Kummer:1994ur}. In \cite{Grumiller:1999rz} its relation to ADM- 
and Bondi-mass as well as its equivalence with the so-called ``mass-aspect 
function'' is discussed. Since the latter has been used in numerical 
calculations as a signal for a black hole \cite{Choptuik:1993jv} we believe 
that the ``BH'' part in ``VBH'' is justified. As to ``virtual'' we would 
like to point out that asymptotically the geometry is equivalent to 
Minkowski spacetime. So a black hole does not exist in the initial or final 
states and appears only at an intermediate stage. This explains the notion 
``virtual''.

In this respect our VBH is quite similar to Hawking's VBH. Also their 
corresponding topologies resemble each other: ours is $\Si\times S^2$, with
$\Si$ being the (compact) light-like cut. Of course, there has to be a 
difference to the Euclidean $S^2\times S^2$ (where the notion of 
``light-like'' does not exist, unless one complexifies) because after all our 
VBH inherits the Minkowski signature. Note that in both cases the first Betti 
number vanishes and the second is non-vanishing. This property was the main 
reason why Hawking considered $S^2\times S^2$ as an interesting building block 
of space-time foam \cite{Hawking:1996ag}.

Finally, let us address the issue of the two endpoints 
($y$ and the intersection of the cut with the $r=0$ line) of the cut, which 
are obviously absent in the Euclidean version: we have already emphasized that 
our calculations have been performed in a specific gauge. In other words, a
peculiar slicing has been introduced. If one would slice an $S^2$ one would 
also obtain two ``endpoints'' corresponding to the intersection points of the
two tangential hyper-planes in the family of parallel hyper-planes slicing the 
2-sphere. This shows that the two notions of VBH are not completely different 
after all\footnote{Unfortunately, there is no straightforward 
``Euclideanization'' of our path integral quantization scheme, since an 
important technical point was the use of light-cone gauge for the Minkowski 
metric. There is no Euclidean equivalent to this gauge. Moreover, Hawking VBHs
are instanton solutions (i.e. a dominating contribution to the Euclidean path 
integral which describes tunneling/pair creation amplitudes) and there is no 
comparable object in Minkowski spacetime.}.

\section{Outlook}

We have shown that virtual black holes (VBHs) provide non-trivial 
phenomenology already in the simple case of the spherically reduced 
Einstein-massless-Klein-Gordon model. They enter the $S$-matrix, an idea put 
forward some time ago in \cite{'tHooft:1996tq}.
Of course, $4d$ gravity -- even in the spherically symmetric case -- is 
more complex and classical equivalence to a $2d$ theory by no means implies 
quantum equivalence since, roughly speaking, renormalization and dimensional
reduction need not commute (``dimensional reduction anomaly'' 
\cite{Frolov:1999an,Sutton:2000gm}). But
since our results were obtained for the low energy limit any UV cut-off 
dependency should be irrelevant for us. So it is likely that similar features
as discussed in the present and previous work \cite{Fischer:2001vz} (forward 
scattering, decay of $s$-waves)
will occur in the full $4d$ theory close to spherical symmetry.

We extended the discussion of \cite{Fischer:2001vz,Grumiller:2001ea} and found 
$CPT$ invariance of the cross-section (\ref{crosssection}). Moreover, we found 
pseudo-self-similar behavior and non-trivial substructure (which breaks that 
self-similarity) in its kinematic sector. Together with the monomial energy 
scaling behavior this might lead towards an easier derivation of the simple 
result (\ref{RESULT}), if one could prove these properties from first 
principles.

We studied the VBH geometry (\ref{ds}) in some detail (including the 
presentation of a Carter-Penrose diagram despite its non-local nature) and 
found distributional contributions to the ensuing Ricci-scalar 
(\ref{ricciscalar}). The corresponding Einstein-Hilbert action (\ref{EH}) was 
found to be vanishing.

Finally, we compared our notion of VBHs with Hawking's Euclidean bubble 
definition \cite{Hawking:1996ag}. We found parallels in the topological 
structure. Still, the two notions are quite different, since our results 
cannot be continued straightforwardly to Euclidean space (there are no 
Minkowskian instantons).

Within our perturbation theory there are two logical next steps: one
could consider additional external scalar fields at tree level or start with 
one-loop calculations. The first route would be straightforward, but 
since already the 4-point vertex at tree level involved quite lengthy 
transformations and little new is to be expected from such a calculation this 
would probably be a misdirection of resources. Moreover, the contribution to 
the amplitude will be suppressed by additional powers of the total energy. The
only relevant issue could be a sizable contribution in a region where our
4-point cross-section was almost vanishing. The loop calculation, however,
seems more interesting, at least as far as the physical content is concerned: 
we would gain further insight into the information paradox in a region, which
is usually not easily accessible to standard black hole physics: microscopic 
black holes (with total mass much smaller than Planck mass). Preliminary 
results look promising \cite{gkvprep}.

It would also be of interest to consider fermions. We expect qualitative 
changes due to the following observations: on the one hand, the constraint 
algebra will be quite different -- we already have witnessed the tremendous 
change between minimal and nonminimal coupling. On the other hand, our 
amplitude (\ref{RESULT}) contains in the language of elementary particle 
physics the sum over all channels ($s$, $t$, and $u$). However, with a 
conserved charge (like fermion number) typically a sum over only two channels
occurs (cf. e.g. \cite{peskinquantum}). Moreover, the dimensional reduction
procedure is more involved for this case since one has to consider the
spin structure. Roughly speaking, fermions ``feel'' the underlying geometry
much more than scalar particles. In particular, terms coupled to the auxiliary
fields $X^\pm$ and a dilaton dependent ``mass'' term will appear in the matter
Lagrangian \cite{hofmann}.
So new results are to be expected from $s$-wave fermion scattering.

A final {\em ceterum censeo}: as long as a 
comprehensive quantum theory of gravity does not exist, $2d$ 
dilaton gravity will remain an active field trying to solve the conceptual 
issues of quantum gravity in a technically more manageable framework.

\vspace{0.5cm}

{\em Note added in proof:}
According to an erratum to \cite{Fischer:2001vz} (which is in the process
of publication) the line element (10) should read
$(ds)^2=2drdu+(1-2m(r,u)/r-a(r,u)r-d(r,u))(du)^2$ with
$d(r,u):=\delta(u-u_0)\theta(r_0-r)d_0(r_0)$. This leads to additional 
contributions to the Ricci scalar (12) which can be calculated with the   
methods described in appendix A. The result is $R_{\rm new}=R_{\rm old}+
\delta(u-u_0)\left[-\delta'(r-r_0)d_0-\delta(r-r_0)4d_0/r+
\theta(r_0-r)2d_0/r^2\right]$. However, these new terms do not change
any of the conclusions of the present work (in particular, the action
(13) still vanishes).

\section*{Acknowledgment}

This work has been supported by project P-14650-TPH of the Austrian Science 
Foundation (FWF). I would like to thank H. Balasin, V. Berezin, D. Hofmann,  
D. Schwarz, R. Wimmer and particularly my collaborators P. Fischer, W. Kummer 
and D. Vassilevich for stimulating discussions. 

\section*{Appendix A - curvature of VBH geometry}

For a primitive discussion of the Ricci-scalar we need only the equation of 
motion following from (\ref{dil}) when varying the dilaton field $X$:
\begin{equation} 
d \wedge \omega - e^- \wedge e^+ \left(V'(X)+X^+X^-
U'(X)\right) + \frac{\delta {\cal{L}}^{(m)}}{\delta X} = 0, \label{eom3}
\end{equation}
The first observation is that the last term $\de {\cal L}^{(m)}/\de X$ 
contains $\de$-like contributions. This follows from equations (8) and (9) of 
\cite{Fischer:2001vz}, which localize the trace of the $4d$ energy-momentum 
tensor (which is proportional to the term $\de {\cal L}^{(m)}/\de X$). The 
second observation concerns the relation between the Ricci-scalar and the 
first term, $d\wedge\om$: they are (apart from numerical pre-factors) Hodge 
dual to each other. This implies that the Ricci-scalar has a $\de$-like 
contribution at the point denoted by $y$ in figure \ref{fig:cp}. The
second term has a discontinuity at the same point, inherited from the 
discontinuity in the metric. The curvature vanishes trivially beside the
cut and it is non-vanishing on the cut. That is why we said the curvature 
singularity at $y$ ``propagates'' along the light-like cut to the origin.
Alas, there is a ``subtlety'' involved in this discussion: one term contains 
the product of two $\de$-functions at the same point, i.e. it is not a 
well-defined distribution. 

Therefore, we will discuss the curvature forgetting about the first order 
formalism and start directly with the line element (\ref{ds}) supplemented
by the angular part $-r^2d^2\Om$. In complete analogy to \cite{Balasin:1994kf} 
we use the Kerr-Schild decomposition $g_{ab}=\eta_{ab}-fk_ak_b$ (note the
relative sign due to differing signature conventions between this work and
\cite{Balasin:1994kf}) with the profile function $f$, a null vector field 
$k_a$ and a flat background metric $\eta_{ab}$. This representation is 
tailor-made for distributional contributions in $f$, since the Ricci scalar 
($k^a=\eta^{ab}k_b=g^{ab}k_b$)
\eq{
R=\partial_a\partial_b\left(fk^ak^b\right)
}{ricci}
contains the profile function linearly and hence problems with ``squared 
$\de$-functions'' will not arise. Using again outgoing Sachs-Bondi coordinates 
$(ds)^2=2drdu+(du)^2-f(du)^2$ the profile function reads
\eq{
f=\left(\frac{2m_0}{r}+a_0r\right)\de(u-u_0)\Theta(r_0-r),
}{f}
and as additional simplifications we have $(k\cdot\partial)=\partial_r$ and
$(\partial\cdot k)=2/r$. Thus, we obtain
\meq{
\left.R\right|_{r>0}=\partial_r\partial_rf + \frac{4}{r}\partial_rf +
\frac{2}{r^2}f = \de(u-u_0)
{\Bigg [}-\de'(r-r_0)\left(\frac{2m_0}{r}+a_0r\right) \\
-\de(r-r_0)\left(\frac{4m_0}{r^2}+6a_0\right)+\Theta(r_0-r)\frac{6a_0}{r}
{\Bigg ]}.
}{R}
According to \cite{Balasin:1994kf} we will obtain in addition a 
$\de$-like contribution at the origin (the Schwarzschild singularity) of the 
form $\left.R\right|_{SS}=\de(u-u_0)\de(r)2m_0/r^2$ with 
$\int_0^\infty dr\de(r):=1$.
Thus, the curvature scalar is really concentrated on the light-like cut 
depicted in figure \ref{fig:cp} with distributional contributions on its 
endpoints and it vanishes for $r>r_0$.  
Note that the only nonvanishing contribution along the cut is due to the 
Rindler term (this follows simply from $\left.R\right|_{r>0}=0$ for 
Schwarzschild geometry).

\bibliographystyle{unsrt}
\bibliography{comments} 

\end{document}